\documentclass[a4paper,12pt]{article}

\usepackage{amsmath}
\usepackage[final]{graphicx}
\usepackage{bm}
\usepackage{amssymb}

\usepackage{amssymb}
\newcounter{comment}

{\refstepcounter{comment}%
\begin{quote}
\ttfamily\small$\blacksquare$ \textbf{\underline{Comment} $\sharp$\thecomment:}}%
{\end{quote}}

\begin{document}
\hfill
\begin{minipage}{20ex}\small
ZAGREB-ZTF-08-02\\
\end{minipage}

\begin{center}
\baselineskip=2\baselineskip
\textbf{\LARGE{
Nondecoupling of a terascale isosinglet quark and rare $K$ and $B$ decays
}}\\[6ex]
\baselineskip=0.5\baselineskip

{\large Ivica~Picek
\footnote{picek@phy.hr;
} and
Branimir~Radov\v{c}i\'c
\footnote{bradov@phy.hr}}\\[4ex]
\begin{flushleft}
\it
Department of Physics, Faculty of Science, University of Zagreb,
 P.O.B. 331, HR-10002 Zagreb, Croatia\\[3ex]
\end{flushleft}
\today \\[5ex]
\end{center}

\begin{abstract}

We examine recent extensions of the standard model with an up-type vectorlike
isosinglet $T$ quark that mixes dominantly with the top quark. We
take under scrutiny the nondecoupling effects which may reveal such a
new heavy fermion through loop diagrams relevant for rare decays such as
$K\to \pi \nu \bar{\nu}$, $B\to \pi (K) \nu \bar{\nu}$, and $B_{s,d}\rightarrow\mu^+\mu^-$. After
demonstrating in detail the cancellation between the leading
nondecoupling terms, we show that two residual forms  $\sim
s^2\ln{m_T^2}$ and $\sim s^4m_T^2$ act in a complementary way, so that
the maximal allowed values of the decay rates are practically
independent of ${m_T}$. While they correspond to $\sim 20\%$ or
$\sim 30\%$ corrections to the SM  rates for $K\to
\pi \nu \bar{\nu}$ and $B\to \pi (K) \nu \bar{\nu}$, an increase by $\sim 50\%$ for
$B_{s,d}\rightarrow\mu^+\mu^-$ decays offers a
possibility to reveal an additional isosinglet state
by measurements of these decays at the Large Hadronic Collider.

\end{abstract}

\vspace*{2 ex}

\begin{flushleft}
\end{flushleft}

\clearpage

\section{Introduction}

Awaiting the forthcoming results from the Large Hadronic Collider (LHC), we are trying to
imagine ways to probe the conceivable new degrees of freedom
belonging to the physics at the TeV scale. In case the mass of
a new degree of freedom lies out of the reach of direct LHC
production, one could still hope to infer on it from its virtual
loop effects. The well-known examples from the past are charm- and
top-quark loop effects in the box diagrams contributing to $K -
\bar{K}$ and $B - \bar{B}$ mixing, respectively. There has been a
recent revival of the latter box diagrams by Vysotsky
\cite{Vysotsky:2006fx}, in the context of  a simple extension of the
standard model (SM) with an additional up-quark state. The essential
reason for examining such an additional degree of freedom has been the
so-called {\em nondecoupling} of a heavy quark in the box diagrams.
When the loop diagrams with exchanged Goldstone boson and heavy
quark are evaluated in the renormalizable gauge, this nondecoupling
is known to originate in the Yukawa coupling proportional to the
heavy quark mass. Generally, in rare processes the box diagrams
combine with the Z-penguin diagrams that exhibit similar
nondecoupling. Therefore, we extend Vysotsky's investigation of the
box diagrams to the Z-penguin diagrams, and explore their impact on
some phenomenologically interesting rare decay processes.

Our present dwelling on the up-quark sector is motivated in part by
a recent study by Alwall et al.\cite{Alwall:2006bx} which showed
that the Tevatron measurements leave ample space for an extra top state
above 256 GeV \cite{CDFRun2}. An extra top state is expected to be
part of some beyond the standard model (BSM), but the results of the
present paper are quite general and do not rely on any specific BSM.
Note that there is even a possibility that such an extra state originates
on account of the nonperturbative effects within the SM itself.
There has been a proposal of such a bound state of 6 top and 5
antitop quarks \cite{Froggatt:2004bh} that could be probed at the
LHC.

Extra vectorlike quark states of both up- and down-type
were already subjected to detailed previous studies
\cite{paconpb,london,barger,largo}. The more recent study that
focused on the down-quark sector \cite{Grossman:2007bd} is
complementary to what we are exploring here in the up-quark sector.
Moreover, being in favor of a rather heavy spectrum of isosinglet
down states, the results by these authors indicate that an isosinglet
up-quark could be well beyond the direct LHC production.

Our starting point is the model advocated by Vysotsky
\cite{Vysotsky:2006fx}, where the new up-type quark is just above
the LHC reach. This model has an appeal of repeating in the up-quark
sector a sort of seesaw mechanism, that is known to be operative in
the up-lepton (neutrino) sector. By imposing the quark-lepton
symmetric generalization of the SM, the existing modification of the
SM in the leptonic upper-hypercharge (neutrino) sector opts for the
corresponding extra up-type isosinglet quark sector, rather than a
full sequential fourth family. Then, a new heavy quark mass scale
$M$ ensures a smallness of the mixing of an additional top quark
with the SM top and may explain why the SM top is much heavier than
other SM fermions \cite{Vysotsky:2006fx}.

In the next section we first describe the predictive two-parameter
model based on  such an isosinglet extension. We first check the
allowed parameter space against the flavor-conserving electroweak
precision observables (EWPO). The obtained constraints are drawn on
account of the nondecoupling, both in the flavor specific parameter
$R_b$ and in the oblique $T$ parameter. In the third section we
focus on the flavor-nondiagonal Z-penguin and box diagram transitions
that exhibit a subtle cancellation of the leading nondecoupling terms.
Finally, we analyze the sensitivity of the golden $K\to \pi \nu \bar{\nu}$, $B\to \pi (K) \nu \bar{\nu}$,
and $B\to \mu^+ \mu^-$ decays to such an extension of the SM and
compare it to the results 
of the littlest
Higgs model (LHM) that provides a monitoring case.

\section{EWPO constraints on model with extra isosinglet quark}

It is conceivable that possible new quark states can be out of the
direct production reach also in the era of the LHC. Therefore, we
take a closer look at the model proposed recently by the authors of
Ref. \cite{Alwall:2006bx} and check that the EWPO constraints allow
for a wider parameter span than considered  in \cite{Alwall:2006bx}.
This motivates us to further extend the approach by Vysotsky
\cite{Vysotsky:2006fx} toward a still unexplored region of the
parameter space and to focus on rare decays which may be sensitive to
this extended parameter region.

The model at hand represents an extension of the SM by the heavy
isosinglet, which in the case of Ref. \cite{Alwall:2006bx} is  a new
up-type $T$-quark with a mass just above the Tevatron bound of 256
GeV \cite{CDFRun2}. Vysotsky considers, instead, a state with
the mass just above $\simeq$ 5 TeV, the limit of the direct production
at the LHC, and in the present paper we further extend the considered mass region.
 This allows us to look at the loop effects for
the whole allowed parameter space, irrespective of whether or not
the new state can be produced directly. In the present paper we
show that  at $m_T \sim 7$ TeV the extended mass region splits into two regimes
that are dominated by two different types of nondecoupling.
However, the new results of the present paper refer to the regime
of the relatively heavy $T$ state.

The adopted model, expressed in terms of the weak (primed)
eigenstates, reads in the form of the Lagrangian that in addition to
the usual SM piece has an additional BSM part consisting of two Dirac
mass terms and one Yukawa term:
\begin{equation}
{\cal L}_{BSM} = M \bar T^\prime_L T^\prime_R
+ \left[
\mu_R \bar T^\prime_L t^\prime_R
+ \frac{\mu_L}{v/\sqrt 2} \left(\bar t^\prime, \bar b^\prime \right)_L \Phi^c \; T_R^\prime \right] + h.c. \;\; .
\label{1}
\end{equation}
Two new heavy $SU(2)_L$ singlet states, $T^\prime_{L}$ and
$T^\prime_{R}$, have the Dirac mass term $M$ that is nondiagonal
because  $T^\prime$ mixes with the $t^\prime$ state. This mixing is
given by two terms in the square brackets: the $\mu_R$ term
describes the mixing of two $SU(2)_L$ singlets, $T_L^\prime$ and
$t_R^\prime$, while the $\mu_L$ term describes mixing of the SM weak
isodoublet with the isosinglet state $T^\prime$. Obviously, by
switching off these $\mu_{L,R}$ terms,
the $t^\prime$ field would become the ordinary $t$ quark, the mass
eigenstate of the SM.

The presence of the $T^\prime-t^\prime$ mixing  \cite{Vysotsky:2006fx, Alwall:2006bx} 
has several effects. First, the mass
matrix should be diagonalized, and for the charged current
couplings, the  SM unitary CKM matrix $\textbf{V}_{3\times3}^{SM}$
has to be enlarged to
$\textbf{V}_{4\times3}$, a generalized $4\times3$ CKM matrix
that is not unitary. The adopted assumption that $T^\prime$  mixes
only with the $t^\prime$ implies that the unitary transformation to
mass eigenstates is a simple rotation parametrized by the single (real)
angle $\theta$.
Accordingly, the generalized Cabibbo-Kobayashi-Maskawa (CKM) matrix entries  in the charged
current couplings are
\begin{eqnarray}
\nonumber
  V_{ti} &=& V_{ti}^{SM}\cos\theta , \\
  V_{Ti} &=& V_{ti}^{SM}\sin\theta .
\label{ckm}
\end{eqnarray}
Second, the $T^\prime-t^\prime$ mixing modifies the neutral current
couplings of $U = (u,c,t,T)$ states
\begin{equation}
    {\cal L}_{NC}=-\frac{g}{2\cos\theta_W}Z_{\mu}(\bar U_L\textbf{V}\textbf{V}^\dag\gamma^\mu
    U_L-2\sin^2\theta_WJ_{em}^\mu).
\end{equation}
This induces the flavor changing neutral current (FCNC) part, given by the nondiagonal terms in
\begin{equation}\label{neutralcurr}
    \textbf{V}\textbf{V}^\dag=\left(
                                \begin{array}{ccc}
                                  \textbf{1}_{2\times2} & 0 & 0 \\
                                  0 & \cos^2\theta & \sin\theta\cos\theta \\
                                  0 & \sin\theta\cos\theta & \sin^2\theta \\
                                \end{array}
                              \right).
\end{equation}
However, the built-in restriction that the heavy isosinglet
$T^\prime$ state mixes only with the SM $t^\prime$ quark, leads to
the very predictive 2-parameter model expressed in terms of the
$T^\prime-t^\prime$ mixing angle $\theta$, and the mass $m_T$ of the
new isosinglet quark.

Let us turn to the constraints on these new parameters that can be
drawn from EWPO tests. Both type of loop effects, the universal
(oblique) $T$ parameter, and nonuniversal (flavor specific)
$R_b$ parameter, are sensitive to nondecoupling of heavy quarks.

\begin{figure}
\centerline{\includegraphics[scale=0.80]{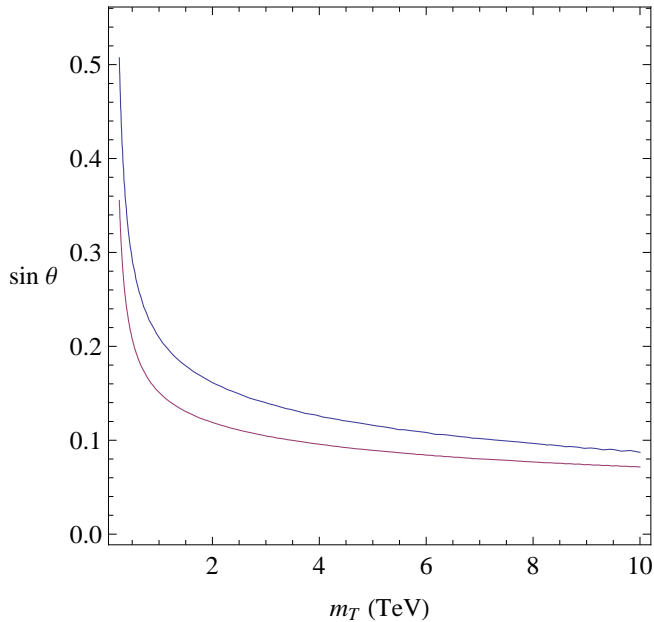}}
\caption{\small Allowed parameter space from EWPO, lying below the upper curve determined 
from $R_b$, or the lower curve determined from the $T$ parameter.}
\label{prostor}
\end{figure}

The effects of the heavy isosinglets in the oblique $T$ parameter  have been
computed in Ref. \cite{Lavoura:1992np}. In the model at hand, where a single isosinglet
quark mixes only with the SM top quark, new contributions to the
$T$ parameter are summarized as
\begin{eqnarray}
\nonumber  T &=& \frac{3}{16\pi\sin^2\theta_W \; \cos^2\theta_W}\{|V_{tb}^{SM}|^2\sin^2\theta \; [\theta_+(y_T,y_b)-\theta_+(y_t,y_b)] \\
   &-& \sin^2\theta  \; \cos^2\theta\ \theta_+(y_T,y_t)\} \; .
\label{T-param}
\end{eqnarray}
Here $y_i=m_i^2/m_Z^2$ and
\begin{equation}
    \theta_+(y_1,y_2)=y_1+y_2-\frac{2y_1y_2}{y_1-y_2}\ln\frac{y_1}{y_2} \; .
\end{equation}
By comparing Eq. (\ref{T-param}) to the most recent experimental value $T=
-0.03\pm0.09$ \cite{Yao:2006px}, we obtain our first constraint on
the allowed part of parameter space $(m_T,\theta)$, lying below the lower curve displayed on
Fig.~\ref{prostor}.

A similar constraint can also be drawn from the flavor specific
parameter $R_b=\Gamma_b/\Gamma_{had}$. Namely, the SM loop diagrams
involving the $t$ quark (with the corresponding parameter
$x_t=m_t^2/m_W^2$) modify the $Zb\bar b$ coupling
\cite{Bernabeu:1990ws}. In particular its left-handed part $g_L^b$
is changed by
\begin{equation}
    \delta g_L^b=\Big(\frac{\alpha}{2\pi}\Big)|V_{tb}^{SM}|^2F(x_t).
\end{equation}
In the BSM at hand, the presence of the $T^\prime-t^\prime$ mixing
modifies this loop correction further,
leading to \cite{Bamert:1996px}
\begin{eqnarray}
\label{tTmodif}
|V_{tb}^{SM}|^2F(x_t)\rightarrow \sum_{j=t,T}|V_{jb}|^2\Big(F(x_j)+
\tilde{F}(x_j)\Big)+V_{tb}^*V_{Tb} \tilde{F}(x_t,x_T) \;.
\end{eqnarray}
The functions $\tilde{F}(x_j)$ and $\tilde{F}(x_t,x_T)$ in Eq.(\ref{tTmodif}) appear also in the flavor changing rare decays, and
they will be presented in more detail in this context in
Eqs. (\ref{f-onR}) and (\ref{f-onC}).
From the expression given in Ref. \cite{Bamert:1996px},
\begin{equation}
    R_b=R_b^{SM}(1-3.56\delta g_L^b+0.645\delta g_R^b+0.00066S-0.0004T)\; ,
\end{equation}
it is clear that the oblique $T$ parameter term can be neglected in the
ratio $R_b$. Using the experimental value $R_b = 0.21629(66)$ and the SM 
prediction $R_b^{SM} =  0.21578(10)$ \cite{Yao:2006px}, 
we obtain our second constraint on the allowed
part of the  parametric space $(m_T,\theta)$. Then, the constraints
obtained from both EW precision parameters can be displayed in the ($m_T,
\sin\theta$) plane, as shown in Fig.~\ref{prostor}. Note that the excluded
parameter space from the universal $T$ parameter suffers from the uncertainties
from the Higgs mass, so that in further we rely on the non-universal parameter $R_b$,
which provides more reliable constraints.

The curves on  Fig.~\ref{prostor} can be compared to previous plots for low $m_T$ values, 
given in \cite{Alwall:2006bx} and \cite{largo}.
Our plots are extended to higher values of $m_T$
\cite{Vysotsky:2006fx},  where the new state
escapes the direct production at the LHC. In the next section we consider possible
effects for the whole allowed parameter space. Thereby we will show how  these effects 
originate in a nondecoupling of the heavy isosinglet quark 
in the loop diagrams for selected rare decays.\\

\section{The structure of nondecoupling and its manifestation in selected rare decays}

The model introduced in the previous section has been applied by
Vysotsky  \cite{Vysotsky:2006fx} to $\Delta F = 2$,  $B - \bar{B}$ mixing, represented by
the quark-box diagrams. Now we extend such a study to the quark-lepton box
diagrams and
Z-penguin diagrams that govern the present day golden decays of
flavor physics, $K^+ \rightarrow\pi^+ \nu\bar\nu$, $K_L
\rightarrow\pi^0 \nu\bar\nu $, $B\to \pi \nu \bar{\nu}$, $B\to K \nu \bar{\nu}$, and $B\rightarrow\mu^-\mu^+$. These FC
transitions are dominated by a nondecoupling existing in the box and
the Z-penguin diagrams. Huge effects are {\em a priori} possible, if there
were no cancellations among the leading nondecoupling terms.
We are left by effective suppression of
nondecoupling that will appear in the SM
Inami-Lim functions \cite{InL} describing the rare decays
$K\rightarrow\pi\nu\bar\nu$, $B\to \pi (K) \nu \bar{\nu}$, and $B\rightarrow\mu^-\mu^+$. These short distance functions, by now, have a standard notation, $X(x_t)$ and $Y(x_t)$
\cite{Buchalla:1990qz}.\\
For the adopted BSM with the extra isosinglet $T$ quark, these
functions are modified by contributions from loops involving this
new quark. In addition to contributions from loops involving only
a $t$ or only a $T$ quark, there are additional contributions from loops
involving simultaneously $t$ and $T$ quarks. In the first case there
is a change in the relevant CKM elements for the charged current
coupling, Eq.~(\ref{ckm}), and a change in flavor diagonal NC
coupling Eq.~(\ref{neutralcurr}), but in the second case also the
FCNC coupling in eq.~(\ref{neutralcurr}) occurs. In general, the
powers of small factor $s\equiv \sin\theta$ will suppress the
nondecoupling of the new heavy $T$ quark. We are going to demonstrate in
detail the structure of modulation of $m_T$-dependent terms by the
powers of $s$.
\begin{figure}
\centerline{\includegraphics[scale=0.9]{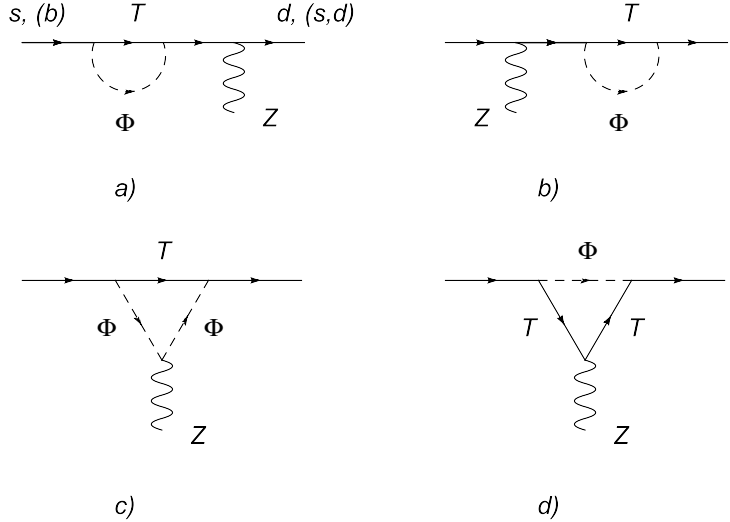}}
\caption{Loop diagrams: a) to c), in which the potentially
leading nondecoupling $\sim s^2x_T$ reduces to merely $\sim s^2\ln x_T$
as demonstrated in
Eq.~(\ref{a+b+h}), and the diagram d) giving the second remaining
form $\sim s^4x_T$.}
\label{iso-penguin}
\end{figure}
The evaluation of relevant Feynman diagrams involving only a $t$ or only
a $T$ quark in the 't Hooft-Feynman gauge has been presented in ref.
\cite{InL}. All divergent and $\sin\theta_W$ dependent terms cancel
mutually in the sum.

Let us first focus on diagrams involving only the $T$ quark. Here,
from the full set of diagrams in Inami and Lim \cite{InL}, we display in Fig.~\ref{iso-penguin} only the most characteristic: The subgroup a) to c) in which the potentially leading nondecoupling $\sim x_T$ cancels out,
and the diagram d) in which such decoupling is suppressed by an additional factor of $s^2$.
In the 't Hooft-Feynman  gauge, the leading nondecoupling  of  the $T$-quark
arises in conjunction with the
Goldstone exchange in off-diagonal self-energy diagrams a) and b), and  in
diagram c). These diagrams, involving only the $T$-quark, will
have a factor of $s^2$ due to the CKM factors in Eq. (\ref{ckm}):
\begin{eqnarray}
\nonumber
  G_{a+b} &\approx& \frac{s^2}{8}\Big(-\frac{1}{2}x_T\ln
    x_T+\frac{3}{4}x_T+{\cal O}(\ln x_T)\Big) \; , \\
  G_c &\approx& \frac{s^2}{8}\Big(\frac{1}{2}x_T\ln
    x_T-\frac{3}{4}x_T+{\cal O}(\ln x_T)
    \Big) \; .
\label{a+b+h}
\end{eqnarray}
The diagram d) on Figure \ref{iso-penguin}, involving NC coupling from Eq. (\ref{neutralcurr}),  will get another factor of $s^2$:
\begin{equation}
 G_{d} \approx \frac{s^4}{8}\Big(x_T-\ln x_T\Big) \; .
\label{only-d}
\end{equation}
The leading nondecoupling terms $\sim s^2 x_T$ and $\sim s^2 x_T \ln
x_T$ in Eq. (\ref{a+b+h}), that are only suppressed by a factor of $s^2$, could {\em a priori} lead to huge effects in FC rare decays. That is not the case
because they cancel mutually in the sum, and only the terms of the
form $\sim s^4 x_T$ and $\sim s^2  \ln x_T$ survive in the sum. 
This cancellation is not accidental, but is known to be a consequence of the SU(2) gauge
symmetry. For an underlying Lagrangian (1), the effective sdZ vertex is generated by the SU(2) breaking in the loops, corresponding to $\mu_L$. The explicated cancellation ensures that
in the limit of large $m_T$ with fixed Yukawa coupling, the $T$ contribution decouples.
Then, after adding together the contributions from all diagrams, we
are left with 
a relic
of nondecoupling of $T$ quark that has the form:
\begin{equation}\label{contribT}
    G(x_T) = s^2P(x_T)+s^4R(x_T) \; .
\end{equation}
Here
\begin{eqnarray}
  P(x) &=& \frac{x}{8(x-1)}\Big(3+\frac{5x-8}{x-1}\ln x\Big) \; , \\
  R(x) &=& \frac{1}{8}\Big(x-2\frac{x\ln{x}}{x-1}\Big) \; .
\end{eqnarray}
From here, the contributions from loops involving only the $t$-quark can be obtained by
simple replacements $s\rightarrow c$ ($c\equiv \cos\theta$)  and $x_T\rightarrow x_t$ in Eq.
(\ref{contribT}):
\begin{equation}
    G(x_t) = c^2P(x_t)+c^4R(x_t) \; .
\end{equation}
Additional Feynman diagrams involving both $t$ and $T$ quarks give
a contribution that also has  only
terms
of the form $\sim s^2 \ln x_T$:
\begin{eqnarray}
\nonumber
  c^2s^2C(x_t,x_T) &=& c^2s^2\frac{1}{4}\Big[\frac{-1}{x_T-x_t}\Big(\frac{x_T^2\ln{x_T}}{x_T-1}-\frac{x_t^2\ln{x_t}}{x_t-1}\Big) \\
   &+& \frac{x_Tx_t}{x_T-x_t}\Big(\frac{x_T\ln{x_T}}{x_T-1}-\frac{x_t\ln{x_t}}{x_t-1}\Big)\Big] \; .
\end{eqnarray}
The functions $R(x)$ and $C(x_t,x_T)$ are related to the functions $\tilde{F}(x_j)$ and $\tilde{F}(x_t,x_T)$ in Eq.(\ref{tTmodif}) by:
\begin{equation}
\label{f-onR}
    \tilde{F}(x_t)=-\frac{1}{\sin^2\theta_W}s^2R(x_t)\; ,\; \tilde{F}(x_T)=-\frac{1}{\sin^2\theta_W}c^2R(x_T)\; ,
\end{equation}
\begin{equation}
\label{f-onC}
    \tilde{F}(x_t,x_T)=\frac{1}{\sin^2\theta_W}s\ c\ C(x_t,x_T)\; .
\end{equation}
After obtaining the upper loop functions for the BSM at hand, we are
ready to present the generalized Inami-Lim functions for specific
rare processes. The  function $X^{SM}(x_t)$ that is relevant for $K\rightarrow\pi\nu\bar\nu$
and $B\to \pi (K) \nu \bar{\nu}$ decays in the SM, in the BSM at hand is
replaced by
\begin{equation}
    X^{SM}(x_t)\rightarrow X^{BSM}=G(x_t)+G(x_T)+c^2s^2C(x_t,x_T) \; .
\end{equation}
By collecting the contributions of the same order in $s^2$ we find
\begin{eqnarray}\label{X-f-on}
\nonumber
  X^{BSM} &=& X^{SM}(x_t)+s^2[-X^{SM}(x_t)-R(x_t)+X(x_T)-R(x_T)+C(x_t,x_T)] \\
   &+& s^4[R(x_t)+R(x_T)-C(x_t,x_T)] \; .
\end{eqnarray}
Similarly, we generalize  the function $Y^{SM}(x_t)$ that is relevant for
the $B\rightarrow\mu^-\mu^+$ decay:
\begin{eqnarray}\label{ybsm}
\nonumber
  Y^{BSM} &=& Y^{SM}(x_t)+s^2[-Y^{SM}(x_t)-R(x_t)+Y(x_T)-R(x_T)+C(x_t,x_T)] \\
   &+& s^4[R(x_t)+R(x_T)-C(x_t,x_T)] \; .
\end{eqnarray}
For  both cases, $X^{BSM}$ and $Y^{BSM}$, the 
parts
that are represented by terms of order $s^2$ are of the form
$s^2\ln{x_T}$ and will be dominant for a relatively light $T$
quark. 
We will show that for a heavy $T$ quark, exceeding $m_T\gtrsim 5$ TeV, the
terms of the form 
$s^4x_T$ will also be important.

\begin{figure}
\centerline{\includegraphics[scale=0.90]{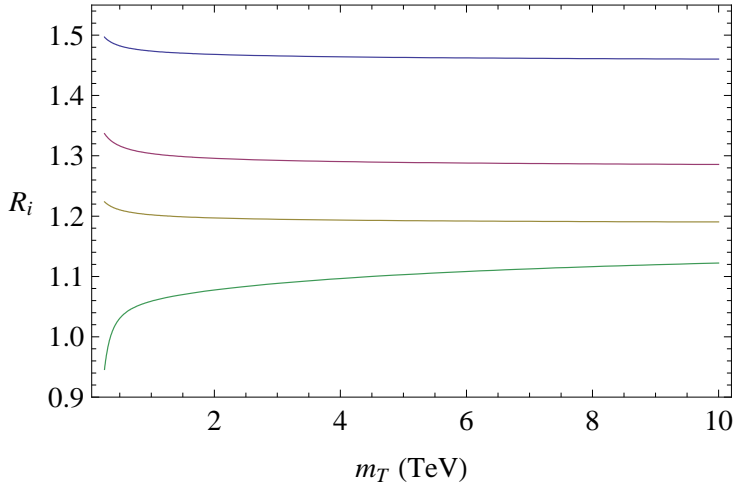}}
\caption{\small Maximal possible enhancements with respect to the SM for the BSM branching ratios  $R_{s,d}$, $R_L$, and $R_+$, going from top down. For comparison we show the corresponding enhancement for the $B - \bar{B}$ mixing, as the lowest curve.}
\label{bratio}
\end{figure}

\begin{figure}
\centerline{\includegraphics[scale=0.90]{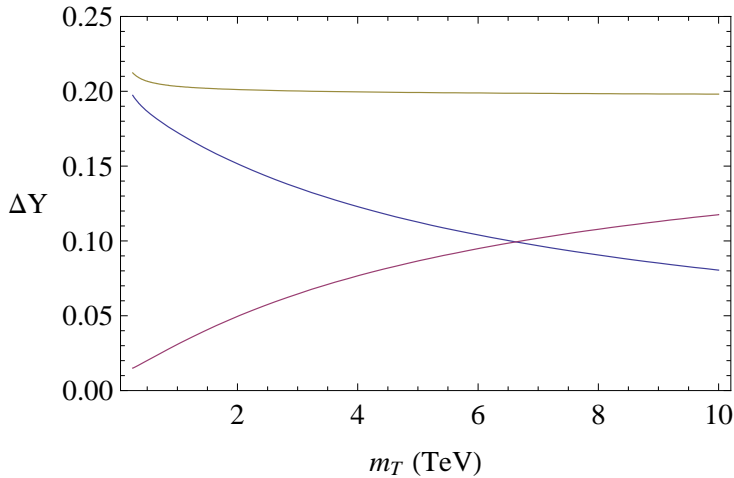}} \caption{\small
The complementary behavior with respect to $m_T$ of the 
two relics of nondecouplings, entering
  $Y^{BSM} = Y^{SM} + \Delta Y$ in Eq. (\ref{ybsm}) relevant for 
$R_{s,d}$: the decreasing
 $\sim s^2$ and the increasing $\sim s^4$ term in Eq. (\ref{ybsm}), and their
almost $m_T$-independent sum on the upper curve.}
\label{terms}
\end{figure}

We will explicate it on the branching ratios for rare decays, which  we define like in Ref.
\cite{Buras:2006wk}:
\begin{eqnarray}
R_+&\equiv&\frac{Br(K^+\rightarrow\pi^+\nu\bar\nu)_{BSM}}{Br(K^+\rightarrow\pi^+\nu\bar\nu)_{SM}} \; , \\
R_L &\equiv& \frac{Br(K_L\rightarrow\pi^0\nu\bar\nu)_{BSM}}{Br(K_L\rightarrow\pi^0\nu\bar\nu)_{SM}}=\frac{Br(B\to\pi(K)\nu\bar{\nu})_{BSM}}{Br(B\to \pi (K) \nu \bar{\nu})_{SM}}=\Big[\frac{X^{BSM}}{X^{SM}}\Big]^2 \;  \\
R_{s,d} &\equiv&
\frac{Br(B_{s,d}\rightarrow\mu^+\mu^-)_{BSM}}{Br(B_{s,d}\rightarrow\mu^+\mu^-)_{SM}}=\Big[\frac{Y^{BSM}}{Y^{SM}}\Big]^2 \; .
\end{eqnarray}
By using constraints from EWPO displayed in Fig.~\ref{prostor} we are able to find maximal allowed values for the ratios $R_+$,
$R_L$ and $R_{s,d}$ as  functions of $m_T$. These plots are  drawn on Fig.~\ref{bratio}.
For comparison, in the same figure we reproduce also the ratio for
the quark-box contribution to the $B - \bar{B}$ mixing considered
by Vysotsky \cite{Vysotsky:2006fx}. In contrast to his box diagram case,
all our ratios are slowly decreasing functions of $m_T$. Typically,
for say $m_T=5$ TeV, these maximal ratios $R_+$, $R_L$ and $R_{s,d}$ reach the values of $\sim$ 1.2, 1.3, and 1.5, respectively. Among them, $R_+$ has the smallest enhancement due to a nonnegligible charm contribution that is
unaffected by $t-T$ mixing. On the other hand, $R_{s,d}$ can reach
values of $\sim1.5$ almost independently of $m_T$. The reason is the
complementary behavior of the 
two relics of 
nondecoupling forms, $s^2\ln{x_T}$ and $s^4x_T$, building Eq.(\ref{ybsm}) and displayed in Fig.~\ref{terms} separately, and then as a sum.
For $m_T \geq 7$ TeV the  $m_T$-dependence
is effectively changed from  a $\sim \ln{m_T^2}$ to
a $\sim m_T^2$ form, 
but 
 with an extra factor of $s^2$. The same pattern applies
to the decays $K\rightarrow\pi\nu\bar\nu$ and $B\to \pi (K) \nu \bar{\nu}$, and we do not repeat
here the plots for them.

\begin{figure}
\centerline{\includegraphics[scale=0.90]{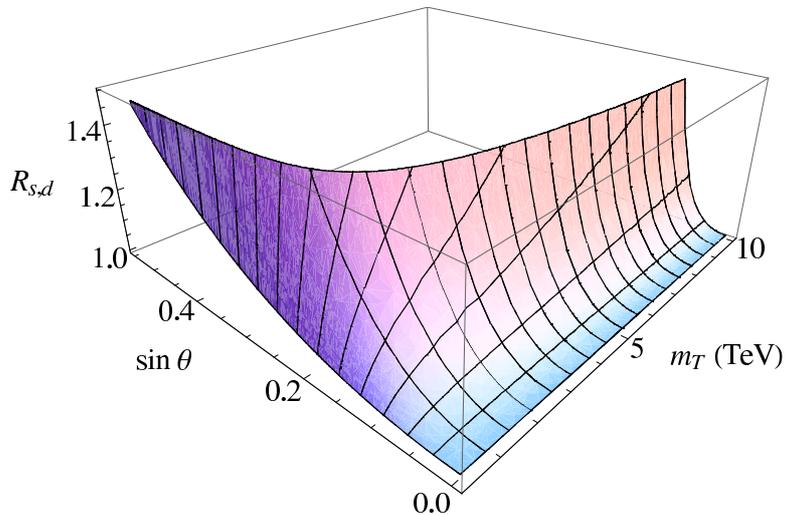}}
\caption{\small Full plot for possible enhancement of the Br($B\rightarrow\mu^-\mu^+$), given by the functional  dependence of the $R_{s,d}$ on the parameters of the model. Let us note that by projecting the top of the displayed surface on the ($R_{s,d}$, $m_T$) plane, one obtains the uppermost
curve in Fig.~\ref{bratio}.}
\label{bmimi}
\end{figure}

The obvious favorite  $B_{s,d}\rightarrow\mu^-\mu^+$ decays are ideally
suited for an investigation at the LHC  (\cite{Fleischer:2008uj} and references therein).
The current upper bounds from the CDF Collaboration read
\begin{eqnarray}
\nonumber
  BR(B_s \rightarrow\mu^-\mu^+) &<& 5.8 \times 10^{-8} , \\
  BR(B_d \rightarrow\mu^-\mu^+) &<& 1.8 \times 10^{-8} ,
\label{BmumuCDF}
\end{eqnarray}
and LHCb will soon reach the possibility to test an increase in $B_{s}\rightarrow\mu^-\mu^+$ 
with respect to the
SM prediction \cite{Blanke:2006ig}:
\begin{eqnarray}\nonumber
\label{BmumuTheo}
  BR(B_s \rightarrow\mu^-\mu^+) &=& (3.35 \pm 0.32) \times 10^{-9} , \\
  BR(B_d \rightarrow\mu^-\mu^+) &=& (1.03 \pm 0.09) \times 10^{-10} .
\end{eqnarray}
Let us therefore display in
Fig.~\ref{bmimi} the $B\rightarrow\mu^-\mu^+$ decay rates as a 3D plot, constrained by
Fig.~\ref{prostor} with
respect to the allowed parameter space. Figure ~\ref{bmimi} illustrates the
possibility to infer on the value of $m_T$ from the measurements of the $B\rightarrow\mu^-\mu^+$ decays, provided that the mixing $\sin\theta$ is fixed from some other observable.

\section{Discussion and Conclusions}

In this paper we examine the loop-induced effects of a possible
terascale up-type isosinglet quark. The quark-lepton symmetry and some additional
arguments listed in the introduction are in favor of such an economical extension of the SM.
We adopt the BSM defined by Eq. (\ref{1}), where the new
$T$ quark modifies the weak and scalar interactions involving
$Q=2/3$ quarks, while the strong and
electromagnetic interactions are not affected. Such a minimal intervention
in the gauge sector satisfies the requirements of minimal flavour violation,
that was recently in focus of Ref. \cite{Grossman:2007bd} for
the down-quark variant of the Lagrangian (\ref{1}). The same study points to a 
spectrum with rather heavy extra down-quark states, that are generally beyond the direct LHC reach. Likewise,
the new up-type quark investigated here is expected to be rather heavy, and is
supposed to be a part of some more complete BSM. A quest for
the complete BSM may require a more general
principle, in a spirit of the Freedberg-Lee symmetry \cite{Friedberg:2007uk} discussed recently by Jarlskog \cite{Jarlskog:2007qy}.

The adopted model represents a predictive
scenario with only two new parameters: the mass of the new $Q=2/3$
isosinglet quark ($m_T$) and its mixing angle to the neighboring
standard model t quark state.
The loop effects of heavy top quarks are governed by the characteristic
forms 
expressed in terms of these two parameters.
The "nondecoupling" of the $T$ contributions holds for increasing $m_T$
with fixed $s$, which in the context of the Lagrangian (1) implies an increasing Yukawa coupling $\mu_L$. However, our priority is to deduce an allowed enhancement of selected SM processes for $T$ quarks beyond the LHC reach, from the mixing angle satisfying phenomenological bounds, rather than to explore in detail in which region of the parameter space the Yukawa coupling $\mu_L/(v/\sqrt{2})$ remains perturbative.
Thus, the existing bounds on the parameters from the electroweak precision measurements allow us to predict the
maximal possible enhancement of selected golden rare decays, respecting  the parameter space allowed by the EWPO. Our novel results refer to a relatively heavy isosinglet quark that is beyond the direct production reach of the LHC.

Possible enhancements for the $\Delta F = 1$ decay rates considered here turn out to be more significant than for previously considered \cite{Vysotsky:2006fx} $B - \bar{B}$ mixing. The effects shown in Fig.~\ref{bratio}
 are
specific for the quark-lepton box and Z-penguin loop amplitudes, 
which  are
exposed in detail in the present paper.
Eventual mild dependence on $m_T$ is determined by  the complementary character of the two 
relics of nondecoupling.
Namely, after the leading nondecoupling terms for the rare decays such as $K\to \pi \nu \bar{\nu}$, $B\to \pi (K) \nu \bar{\nu}$, or $B\to \mu \bar{\mu}$ cancel out,
we are left with two residual terms that
contribute complementary along the whole range of the  $m_T$ values.
We illustrate this in detail in Fig.~\ref{terms}  for the example of $R_{s,d}$ which,  owing to a particular loop structure of the $B\to \mu
\bar{\mu}$ decays, acquires the largest relative contribution from the
new heavy quark state.

We can compare our results with those of some existing elaborated framework
possessing an extra top quark,
like the littlest Higgs model \cite{ArkaniHamed:2002qy}.
However, this framework contains in addition to an extra
heavy top also new vector bosons and an additional weak-triplet
scalar field. The latter two produce already the tree-level
corrections to EWPO and create the tensions that can  be cured by introducing
T-parity. Since this
requires extra mirror fermions, we limit ourselves to the original LHM without T-parity, elaborated in Ref.\cite{Han:2003wu} and applied by Buras and collaborators \cite{Buras:2006wk}
to rare decays which we are studying here. Truncated further to the sector of the top quarks, this LHM version matches our model and serves as a monitoring case for our
calculations.

\begin{figure}
\centerline{\includegraphics[scale=0.80]{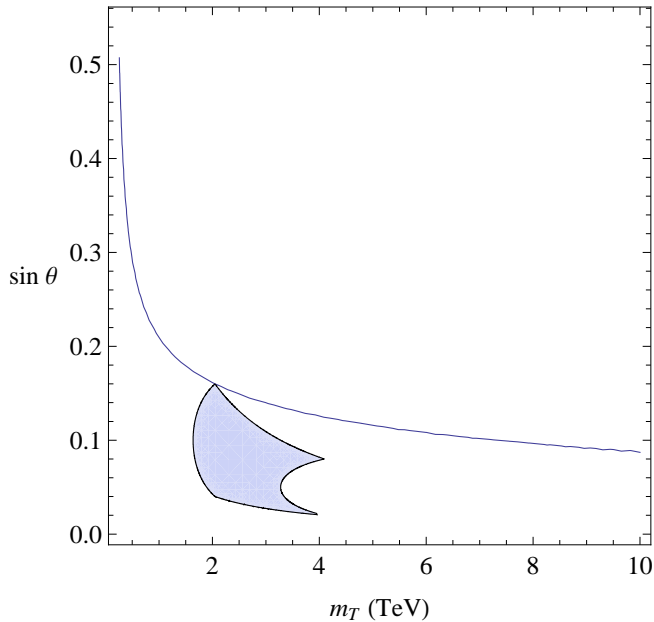}}
\caption{\small LHM parameter region, as restricted by Buras et al., touches the
EWPO curve from Fig.~\ref{prostor} in the point ($s=0.16, m_T=2$ TeV). }
\label{LHMregion}
\end{figure}

Let us note that in the framework with single real mixing angle, CP conserving and CP violating amplitudes are equally enhanced. Extra CP phases enter by allowing the mixing of $T$ to lighter quarks by introducing further nondiagonal Yukawa terms, as attempted recently by Ref. \cite{Fajfer:2007vk} in the context of the LHM. Previously, by such mixing
to extra states, Ref. \cite{largo} achieved large enhancement of  $K_L \rightarrow
\pi^0 \nu \nu$ decays.

Two parameters of our model, $m_T$ and $\sin\theta$, have two corresponding
LHM parameters, $f$ and $x_L$, and can be expressed in terms of them:
\begin{equation}
\label{LHMparam}
    m_T=\frac{f}{v}\frac{m_t}{\sqrt{x_L(1-x_L)}} \; \; , \; \;\sin \theta
=x_L\frac{v}{f}\; .
\end{equation}
Here, $f$ is the new scale of the LHM and $x_L$ is given by a specific ratio of the Yukawa couplings in the LHM sector of top quarks.
New contributions to the functions $X$ and $Y$ of the order $s^2$ given in Eqs. (\ref{X-f-on}) and  (\ref{ybsm})
are identical to the contributions of the Feynman diagrams in Fig. 3 in ref. \cite{Buras:2006wk} at the order $x_L^2\frac{v^2}{f^2}$, as given in Eqs. (4.19) and  (4.25) in \cite{Buras:2006wk}. Note that the parameters of the LHM are
constrained in Ref. \cite{Buras:2006wk} to the range $0.2 \leq x_L \leq 0.8$, and 
 $5 \leq f/v \leq 10$.
Mapped to the ($m_T, \sin\theta$) plane, these constraints result in
the shaded region shown in Fig.~\ref{LHMregion}.

Let us stress that the point ($f=5, \; x_L=0.8$), equivalent to ($s=0.16, \; m_T=2$ TeV)
saturates the EWPO bound and corresponds to the extreme values
for the ratios $R_+, R_L,  R_{s,d}$ that can be identified  in Fig.7 in Ref.\cite{Buras:2006wk}.
This touching in a single point with the EWPO curve in Fig.~\ref{prostor} compares to the maximal enhancement  which is  obtained in our model for a much wider region of the parameter space shown in Fig.~\ref{bmimi}. Accordingly, our analysis shows that a rather heavy $m_T$ ($m_T \geq 5$ Tev) has the best chance to be  recovered via an
increase in $B_{s}\rightarrow\mu^+\mu^-$ decay rate measured at the LHC.

\subsubsection*{Note added}
After this work had been completed, we noticed paper \cite{Kopnin:2008ca}
in which the studies  of Ref.\cite{Vysotsky:2006fx} were extended
to rare decays (but not $B_{s,d}\rightarrow\mu^+\mu^-$ decays). In the case of decays governed by the $X^{BSM}$ function, on which
Ref. \cite{Kopnin:2008ca} restricts, we agree with their results.

\subsubsection*{Acknowledgment}
This work is supported by the 
Croatian Ministry  of Science, Education and Sports under Contract No. 119-0982930-1016.
I.~P. gratefully acknowledges Svjetlana Fajfer for bringing to his attention the up-quark sector
in the LHM and for the hospitality offered at the Josef Stefan Institute.
He also acknowledges the stimulating atmosphere of the CERN workshops
``Flavour in the era of the LHC''.


\begin{thebibliography}{10}

\bibitem{Vysotsky:2006fx}
M. I.Vysotsky,,
{\em New (virtual) physics in the era of the LHC,}
Phys. Lett.
{\bf B 644} (2007) 352 [{\tt hep-ph/0610368}].

\bibitem{Alwall:2006bx}
J. Alwall, R. Frederix, J.-M. Gerard, A. Giammanco, M. Herquet, S. Kalinin, E. Kou, V. Lemaitre, F. Maltoni,
{\em Is V(tb) ~= 1?,}
Eur. Phys. J.
{\bf C 49} (2007) 791 [{\tt hep-ph/0607115}].

\bibitem{CDFRun2}
CDF Collaboration,\\
http://www-cdf.fnal.gov/physics/new/top/2005/ljets/tprime/gen6/public.html

\bibitem{Froggatt:2004bh}
C. D. Froggatt, H. B. Nielsen, L. V. Laperashvili,
{\em Hierarchy-problem and a bound state of 6 t and 6 anti-t,}
Int. J. Mod. Phys.
{\bf A20} (2005) 1268 [{\tt hep-ph/0406110}].


\bibitem{paconpb}
F. del Aguila and M. J. Bowick,
{\em The Possibility Of New Fermions With $\Delta I = 0$ Mass,}
Nucl. Phys. {\bf B 224} (1983) 107.

\bibitem{london}
P. Langacker and D. London,
{\em Mixing Between Ordinary And Exotic Fermions,}
Phys. Rev. {\bf D 38} (1988) 886.

\bibitem{barger}
V. D. Barger, M. S. Berger and R. J. N. Phillips,
{\em Quark singlets: Implications and constraints,}
Phys. Rev. {\bf D 52} (1995) 1663 [{\tt hep-ph/9503204}].

\bibitem{largo}
J. A. Aguilar-Saavedra,
{\em Effects of mixing with quark singlets,}
 Phys. Rev. {\bf D 67} (2003) 035003
[{\em Erratum-ibid.} {\bf D 69} (2004) 099901] [{\tt hep-ph/0210112}].




\bibitem{Grossman:2007bd}
Y. Grossman, Y. Nir, J. Thaler, T. Volansky and J. Zupan,
{\em Probing Minimal Flavor Violation at the LHC,}
 Phys. Rev. {\bf D 76} (2007) 096006 [{\tt 0706.1845 [hep-ph]}].

\bibitem{Lavoura:1992np}
L. Lavoura and J. P. Silva,
{\em The Oblique corrections from vector - like singlet and
doublet quarks,}
Phys. Rev. {\bf D 47} (1993) 2046.

\bibitem{Yao:2006px}
W. M. Yao et al. (PDG),
{\em Review of particle physics,}
J. Phys. {\bf G33} (2006) 1.


\bibitem{Bernabeu:1990ws}
J. Bernabeu, A. Pich and  A. Santamaria,
{\em Top quark mass from radiative corrections to the Z $\to$ banti-b decay, }
Nucl. Phys. {\bf B363} (1991) 326.

\bibitem{Bamert:1996px}
P. Bamert, C. P. Burgess, J. M. Cline, D. and  London, and E. Nardi,
{\em $R_b$ and New Physics: A Comprehensive Analysis,}
Phys. Rev.
{\bf D54} (1996) 4275 [{\tt hep-ph/9602438}].

\bibitem{InL}
  T.~Inami and C.~S.~Lim,
 {\em Effects Of Superheavy Quarks And Leptons In Low-Energy Weak Processes $K(L)
 \to \mu \bar{\mu}$, $K^+ \to \pi \nu \bar{\nu}$ and $K^0-\bar{K^0}$,}
  Prog.\ Theor.\ Phys.\  {\bf  65} (1981) 297
  [Erratum-ibid.\  {\bf 65} (1981) 1772].

\bibitem{Buchalla:1990qz}
G. Buchalla, A. J. Buras and M. K. Harlander,
{\em Penguin box expansion: Flavor changing neutral current
 processes and a heavy top quark},
Nucl. Phys. {\bf B349} (1991) 1.


\bibitem{Fleischer:2008uj}
R. Fleischer, 
{\em Flavour Physics and CP Violation: Expecting the LHC}
[{\tt 0802.2882 [hep-ph]}].

\bibitem{Blanke:2006ig}
M. Blanke, A, Buras, D. Guadagnoli
                  and C. Tarantino,
{\em Minimal Flavour Violation Waiting for Precise Measurements
                  of $\Delta M_s, S_{\psi \phi}, A^s_{SL}, |V_{ub}|, \gamma$ and
                  $B^0_{s,d} \to \mu^+ \mu^-$},
JHEP {\bf 0610}:003, 2006
[{\tt hep-ph/0604057}].

\bibitem{Friedberg:2007uk}
R. Friedberg and T. D. Lee,
{\em Hidden Symmetry of the CKM and Neutrino Mapping
                 Matrices}
[{\tt 0705.4156 [hep-ph]}].

\bibitem{Jarlskog:2007qy}
C. Jarlskog,
{\em Neutrino Sector with Majorana Mass Terms and Friedberg-Lee}
[{\tt 0712.0903 [hep-ph]}].

\bibitem{ArkaniHamed:2002qy}
N. Arkani-Hamed, A. G. Cohen, E. Katz and A. E. Nelson,
{\em The littlest Higgs},
JHEP {\bf 07}:034, 2002
[{\tt hep-ph/0206021}].

\bibitem{Han:2003wu}
T. Han, Tao, H. E. Logan, B. McElrath and L-T Wang,
{\em Phenomenology of the little Higgs model, Phys. Rev.}
{\bf D67} (2003) 095004 [{\tt hep-ph/0301040}].

\bibitem{Buras:2006wk}
A. J. Buras, A. Poschenrieder, S. Uhlig, and W. A. Bardeen,
{\em Rare K and B decays in the littlest Higgs model without T-
                 parity},
JHEP {\bf 0611}:062, 2006
[{\tt hep-ph/0607189}].

\bibitem{Fajfer:2007vk}
S. Fajfer and J. F. Kamenik,
{\em On the flavor structure of the littlest Higgs model}
JHEP {\bf 12}:074, 2007
[{\tt 0710.4293 [hep-ph]}].

\bibitem{Kopnin:2008ca}
P. N. Kopnin  and M. I. Vysotsky,
{\em Manifestation of a singlet heavy up-type quark in the
                 branching ratios of rare decays $K \to \pi \nu \bar{\nu}$, $B
                 \to \pi \nu \bar{\nu}$ and $B \to K \nu \bar{\nu}$}
[{\tt 0804.0912 [hep-ph]}].

\end{thebibliography}
\end{document}